\documentclass[twocolumn,english,aps,prb,showpacs,superscriptaddress,floats,
amsmath,amssymb,floatfix,dblfloatfix,nobalancelastpage]{revtex4-1}
\usepackage[T1]{fontenc}
\usepackage[latin9]{inputenc}
\usepackage{color}
\usepackage{amsmath}
\usepackage{amssymb}
\usepackage{graphicx}

\makeatletter
\usepackage{babel}

\usepackage{color}

\makeatother

\usepackage{babel}
\begin{document}
\title{Finite-momentum energy dynamics in a Kitaev magnet}
\author{Alexandros Metavitsiadis}
\email{a.metavitsiadis@tu-bs.de}

\affiliation{Institute for Theoretical Physics, Technical University Braunschweig,
D-38106 Braunschweig, Germany}
\author{Wolfram Brenig}
\email{w.brenig@tu-bs.de}

\affiliation{Institute for Theoretical Physics, Technical University Braunschweig,
D-38106 Braunschweig, Germany}
\begin{abstract}
We study the energy-density dynamics at finite momentum of the
two-dimensional Kitaev spin-model on the honeycomb lattice. Due to
fractionalization of magnetic moments, the energy relaxation occurs through
mobile Majorana matter, coupled to a static $\mathbb{Z}_{2}$ gauge field. At
finite temperatures, the $\mathbb{Z}_{2}$ flux excitations act as an emergent
disorder, which strongly affects the energy dynamics. We show that
sufficiently far above the flux proliferation temperature, but not yet in the
classical regime, gauge disorder modifies the coherent low-temperature
energy-density dynamics into a form which is almost diffusive, with
hydrodynamic momentum scaling of a diffusion-kernel, which however remains
retarded, primarily due to the presence of two distinct relaxation channels
of particle-hole and particle-particle nature. Relations to thermal
conductivity are clarified.  Our analysis is based on complementary
calculations in the low-temperature homogeneous gauge and a mean-field
treatment of thermal gauge fluctuations, valid at intermediate and high
temperatures.
\end{abstract}
\maketitle

\section{Introduction}

Ever since the discovery of large magnetic heat transport in quasi
one-dimensional (1D) local-moment systems \cite{Sologubenko2000a,
Kawamata2008, Hlubek2010, Sologubenko2000b, Hess2001}, the dynamics of energy
in quantum magnets \cite{Zotos1997, Heidrich-Meisner2003,
Heidrich-Meisner2004}, has been a topic of great interest \cite{Hess2007,
FHM2007, Hess2019, Bertini2020}.  Unfortunately, rigorous theoretical
progress has essentially remained confined to 1D \cite{Bertini2020}. Above
1D, understanding energy dynamics in quantum magnets remains an open issue at
large. If magnetic long range order (LRO) is present and magnons form a
reliable quasi particle basis, various insights have been gained for
antiferromagnets and cuprates \cite{Dixon1980, Hess2003, Bayrakci2013,
Chernyshev2016}.  If LRO is absent, and in particular in quantum spin liquids
(QSL) \cite{Balents2010, Balents2016}, energy transport has recently come
into focus as a probe of potentially exotic elementary excitations. In fact,
experiments in several quantum disordered, frustrated spin systems in $D \geq
2$ suggest unconventional magnetic energy dynamics. For {\em bulk} transport,
this pertains, e.g., to quasi 2D triangular organic salts
\cite{Yamashita2010,Bourgeois2019,Ni2019}, or to 3D quantum spin ice
materials \cite{Kolland2012,Toews2012,Tokiwa2018}.  For {\em boundary}
transport, i.e., the magnetic thermal Hall effect, recent examples include
Kagomé magnets \cite{Hirschberger2015a, Watanabe2016, Yamashita2019} and spin
ice \cite{Hirschberger2015b}. A microscopic description of such observations
is mostly lacking.

In this context, Kitaev's compass exchange Hamiltonian on the honeycomb
lattice is of particular interest, as it is one of the few models, in which a
$\mathbb{Z}_{2}$ QSL can exactly be shown to exist \cite{Kitaev2006}.  The
spin degrees of freedom of this model fractionalize in terms of mobile
Majorana fermions coupled to a static $\mathbb{Z}_{2}$ gauge field
\cite{Kitaev2006, Feng2007, Chen2008, Nussinov2009, Mandal2012}. From a
material's perspective, $\alpha$-RuCl$_{3}$ \cite{Plumb2014} may be a
promising candidate for this model \cite{Trebst2017}. Free mobile Majorana
fermions have been invoked to interpret ubiquitous unconventional continua in
spectroscopies on various Kitaev materials, like inelastic neutron
\cite{Banerjee2016, Banerjee2016a, Banerjee2018} and Raman scattering
\cite{Knolle2014, Wulferding2020}, as well as local resonance probes
\cite{Baek2017, Zheng2017}.

Clearly, Majorana fermions should also play a role in the energy dynamics in
putative Kitaev materials. However, bulk thermal conductivity,
i.e., $\kappa_{xx}$ in $\alpha$-RuCl$_{3}$ \cite{Hirobe2017, Leahy2017,
Hentrich2018, Yu2018}, seems to be governed primarily by phonons, with
phonon-Majorana scattering as a potential dissipation mechanism
\cite{Hentrich2018, Yu2018, Metavitsiadis2020a}.  Yet, Majorana fermions may
have been observed in the transverse energy conductivity $\kappa_{xy}$ in
magnetic fields, i.e., in the thermal Hall effect, and its alleged
quantization \cite{Kasahara2018, Yokoi2020}.  In view of this, energy
transport in Kitaev spin liquids has been the subject of several theoretical
studies \cite{Metavitsiadis2017a, Nasu2017, Metavitsiadis2017, Pidatella2019,
Metavitsiadis2019, Metavitsiadis2020b}.  Previous investigations however, have
focused on energy-\emph{current} correlation functions and at {\em zero
momentum} only.

In this work, we take a different perspective and consider the
energy-\emph{density} correlation function directly and at finite
momentum. In particular we will be interested in the role, which the emergent
disorder introduced by thermally excited $\mathbb{Z}_{2}$ gauge field plays
in the long wave-length regime. Therefore, we map out the energy diffusion
kernel of the Kitaev model and its momentum and energy dependence, ranging
from low, up to intermediate temperatures and we contrast this with
expectation for simple diffusion in random systems. The paper is organized as
follows. In Section \ref{sec:mpc}, we briefly recapitulate the Kitaev
model. In Section \ref{sec:rr}, details of our calculations are provided, for
the homogeneous and the random gauge state in Subsections \ref{subsec:homg}
and \ref{subsec:rand}, respectively, and the extraction of the diffusion
kernel is described in Subsection \ref{subsec:Dqw}. We discuss our results in
Section \ref{sec:Results} and summarize in Section \ref{sec:Summary}. An
Appendix, App.~\ref{app1}, on generalized Einstein relations is included.

\section{Model\label{sec:mpc}}

We consider the Kitaev spin-model on the two dimensional honeycomb lattice
\begin{equation}
H=\sum_{{\bf l},\alpha}J_{\alpha}S_{{\bf
l}}^{\alpha}S_{{\bf l}+ {\bf r}_{\alpha}}^{\alpha}\,,\label{eq:1}
\end{equation}
where ${\bf l}=n_{1}{\bf R}_{1}+n_{2}{\bf R}_{2}$ runs over the sites of the
triangular lattice with ${\bf R}_{1[2]}=(1,0), \,
[(\frac{1}{2},\frac{\sqrt{3}}{2})]$ for lattice constant $a{\equiv}1$,
and ${\bf r}_{\alpha=x,y,z} =
(\frac{1}{2}, \frac{1}{2\sqrt{3}}),$ $(-\frac{1}{2},\frac{1}{2\sqrt{3}})$,
$(0,-\frac{1}{\sqrt{3}})$ refer to the basis sites $\alpha=x,y,z$,
tricoordinated to each lattice site of the honeycomb lattice. As extensive
literature, rooted in Ref. \cite{Kitaev2006}, has clarified, Eq. (\ref{eq:1})
can be mapped onto a bilinear form of Majorana fermions in the presence of a
static $\mathbb{Z}_{2}$ gauge $\eta_{{\bf l}}=\pm1$, residing on, e.g., the
$\alpha=z$ bonds
\begin{equation}
H=-\frac{i}{2}\sum_{{\bf l},\alpha}J_{\alpha}
\eta_{{\bf l},\alpha}\,a_{{\bf l}}c_{{\bf l}+{\bf r}_{\alpha}}
\,.\label{eq:2}
\end{equation}
Here we introduce $\eta_{{\bf l},\alpha}$ to unify the notation, with
$\eta_{{\bf l},x(y)}=1$ and $\eta_{{\bf l},z}=\eta_{{\bf l}}$.  There are two
types of Majorana particles, corresponding to the two basis sites. We chose
to normalize them as $\{a_{{\bf l}}, a_{{\bf l}'} \} = \delta_{{\bf l}, {\bf
l}'}$, $\{c_{{\bf m}}, c_{{\bf m}'}\} = \delta_{{\bf m}, {\bf m}'}$, and
$\{a_{{\bf l}}, c_{{\bf m}}\}=0$.  For each gauge sector $\{\eta_{{\bf
l}}\}$, Eq.~(\ref{eq:2}) represents a spin liquid.

For the purpose of this work a local energy density $h_{{\bf c}}$ on some
repeating ``unit'' cluster ${\bf c}$ has to be chosen. Obviously $H=
\sum_{{\bf c}} h_{{\bf c}}$.  As for any local density, the latter does not
fix a unique expression for $h_{{\bf c}}$. Different shapes of the real-space
clusters ${\bf c}$ supporting $h_{{\bf c}}$ will typically lead to differing
high-frequency and short wave-length spectra for its autocorrelation
function. However the low-frequency, long-wave-length dynamics is governed by
energy conservation and will not depend on a particular choice of $h_{{\bf
c}}$ qualitatively \cite{Metavitsiadis2020b}.  For the remainder of this work
we therefore set ${\bf c}={\bf l}$, focusing on $h_{{\bf l}} = -(i/2)
\sum_{\alpha} J_{\alpha} \eta_{{\bf l},\alpha} \, a_{{\bf l}}c_{{\bf l}+{\bf
r}_{\alpha}}$, i.e. the energy density formed by the tricoordinated bonds
around each site on the triangular lattice. Its Fourier transform is $h_{{\bf
q}}=\sum_{{\bf l}}e^{i{\bf q}\cdot{\bf l}}h_{{\bf l}}$ with $h_{{\bf
q}}^{\dagger}=h_{-{\bf q}}$ and $h_{{\bf 0}}=H$

\section{Energy Susceptibilty\label{sec:rr}}

In this section, we present our evaluation of the dynamical energy
susceptibility. We focus on two temperature regimes, namely $T \lesssim
(\gtrsim) T^{\star}$.  Here $T^{\star}$ is the so called flux proliferation
temperature.  In the vicinity of this temperature the gauge field and
therefore fluxes get thermally excited. Previous analysis \cite{Nasu2015,
Metavitsiadis2017, Pidatella2019, Metavitsiadis2020b} has shown, that the
temperature range over which a \emph{complete} proliferation of fluxes occurs
is confined to a rather narrow region, less than a decade centered around
$T^{\star} \approx 0.012J$ for isotropic exchange, $J_{z,y,x}$=$J$,
used in this work, and decrease rapidly with anisotropy \cite{Nasu2015,
Pidatella2019}. Our strategy therefore is to consider a homogeneous ground
state gauge, i.e., $\eta_{{\bf l}}=1$ for $T\lesssim T^{\star}$ and a
completely random-gauge states for $T\gtrsim T^{\star}$. This approach has
proven to work very well on a \emph{quantitative} level in several studies of
the thermal conductivity of Kitaev models \cite{Metavitsiadis2017,
Pidatella2019, Metavitsiadis2017a}.  Note that for $\alpha$-RuCl$_{3}$, the
case of $T\lesssim T^{\star}$ is more of conceptual, than of practical
interest, since it refers to very low temperatures, assuming a generally
accepted $|J|\sim85K$ \cite{Banerjee2016}.

\subsection{Homogeneous gauge for $T\lesssim T^{\star}$\label{subsec:homg}}

For $\eta_{{\bf l}}=1$, the Hamiltonian (\ref{eq:1}) can be diagonalized
\emph{analytically} in terms of complex Dirac fermions. Mapping from the real
Majorana fermions to the latter can be achieved in various ways, all of which
require some type of linear combination of real fermions in order to form
complex ones. Here we do the latter by using Fourier transformed Majorana
particles, $a_{{\bf k}}^{ \phantom{ \dagger}} = \sum_{{\bf l}} e^{-i{\bf k}
\cdot{\bf l}} a_{{\bf l}} / \sqrt{N}$ with momentum ${\bf k}$ and analogously
for $c_{{\bf k}}^{\phantom{\dagger}}$. The momentum space quantization is
chosen explicitly to comprise $\pm{\bf k}$ for each $|{\bf k}|$. Other
approaches, involving reshaped lattice structures \cite{Feng2007,
Nussinov2009}, may pose issues regarding the discrete rotational symmetry of
the susceptibility.

The fermions introduced in momentum space are complex with $a_{{\bf k}}^{
\dagger} = a_{-{\bf k}}^{ \phantom{\dagger}}$, i.e., with only half of the momentum
states being independent. This encodes, that for each Dirac fermion, there
are two Majorana particles. Standard anticommutation relations apply,
$\{a_{{\bf k}}^{\phantom{\dagger}},a_{{\bf k}'}^{\dagger}\}=\delta_{{\bf
k},{\bf k}'}$, $\{c_{{\bf k}}^{\phantom{\dagger}},c_{{\bf k}'}^{
\dagger}\}=\delta_{{\bf k},{\bf k}'}$, and $\{a_{{\bf k}}^{ (\dagger)}
c_{{\bf k}'}^{ (\dagger)}\}=0$. From this, the diagonal form of $H$ reads
\begin{equation}
H=\sum_{{\bf k},\gamma=1,2}^{\sim}\mathrm{sg}_{\gamma}\,\epsilon_{{\bf k}}
\,d_{{\bf k},\gamma}^{\dagger}
d_{{\bf k},\gamma}^{\phantom{\dagger}}\,,
\label{eq:3}
\end{equation}
where the $\tilde{\sum}$ sums over a reduced ``positive'' half of momentum
space and $\mathrm{sg}_{\gamma}$=1(-1) for $\gamma$=1(2). The quasiparticle
energy is $\epsilon_{{\bf k}} = J[3 + 2 \cos(k_{x}) + 4 \cos(k_{x}/2) \cos(
\sqrt{3}k_{y}/2)]^{1/2} /2$.  In terms of reciprocal lattice coordinates
$x,y\in[0,2\pi]$, this reads $\epsilon_{{\bf k}}=J[ 3 + 2 \cos(x) + 2 \cos(y)
+ 2 \cos(x-y)]^{1/2}/2$ with ${\bf k}=x\,{\bf G}_{1}+y\,{\bf G}_{2}$, where
${\bf G}_{1[2]}=(1,-\frac{1}{\sqrt{3}}),\,[(0,\frac{2}{\sqrt{3}})]$.  The
quasiparticles are given by
\begin{align}
\left[\begin{array}{c}
c_{{\bf k}}\\
a_{{\bf k}}
\end{array}\right] & =\left[\begin{array}{cc}
u_{11}({\bf k}) & u_{12}({\bf k})\\
u_{21}({\bf k}) & u_{22}({\bf k})
\end{array}\right]\left[\begin{array}{c}
d_{1{\bf k}}\\
d_{2{\bf k}}
\end{array}\right]\,,\label{eq:4}\\
u_{11}({\bf k}) & =-u_{12}({\bf k})=
\frac{i\sum_{\alpha}e^{-i{\bf k}
\cdot{\bf r}_{\alpha}}}{2^{3/2}\epsilon_{{\bf k}}}\,,\nonumber \\
u_{21}({\bf k}) & =u_{22}({\bf k})=\frac{1}{\sqrt{2}}\,.\nonumber 
\end{align}
From the sign change of the quasiparticle energy between bands $\gamma$=1,2
in Eq. (\ref{eq:3}) it is clear that the relations $a_{{\bf
k}}^{\dagger}=a_{-{\bf k}}^{\phantom{\dagger}}$ and $c_{{\bf
k}}^{\dagger}=c_{-{\bf k}}^{\phantom{\dagger}}$ for reversing momenta of the
original Majorana fermions has to change into $d_{1(2){\bf
k}}^{\dagger}=d_{2(1)-{\bf k}}^{\phantom{\dagger}}$, switching also the
bands. Indeed this is also born out of the transformation (\ref{eq:4}).
Inserting the latter into $h_{{\bf q}}$, the energy density in the
quasiparticle basis reads
\begin{align}
h_{{\bf q}}= & \frac{1}{2}\tilde{\sum_{{\bf k}}}\left\{
\left[\begin{array}{cc}
d_{1{\bf k}+{\bf q}}^{\dagger} &
d_{2{\bf k}+{\bf q}}^{\dagger}
\end{array}\right]\right.\nonumber \\
 & \times\left.\left[
\begin{array}{cc}
\epsilon_{{\bf k}+{\bf q}}{+}\epsilon_{{\bf k}} &
\epsilon_{{\bf k}+{\bf q}}{-}\epsilon_{{\bf k}}\\
\epsilon_{{\bf k}}{-}\epsilon_{{\bf k}+{\bf q}} &
{-}\epsilon_{{\bf k}+{\bf q}}{-}\epsilon_{{\bf k}}
\end{array}\right]\left[\begin{array}{c}
d_{1{\bf k}}^{\phantom{\dagger}}\\
d_{2{\bf k}}^{\phantom{\dagger}}
\end{array}\right]\right\}\,. \label{eq:5}
\end{align}
As to be expected, $h_{{\bf q}{=}{\bf 0}}=H$ from (\ref{eq:3}) and the
off-diagonal interband transitions vanish in that limit.

The energy density susceptibility $\chi({\bf q},z)$ is obtained from Fourier
transformation of the imaginary time density Green's function $\chi({\bf
q},z)=\int_{0}^{\beta}d\tau\langle T_{\tau}(h_{{\bf q}}(\tau)h_{-{\bf
q}})\rangle e^{i\omega_{n}\tau}/T$ by analytic continuation of the Bose
Matsubara frequency $i\omega_{n}=i2\pi nT\rightarrow z\in\mathbb{C}$ and
eventually $z\rightarrow\omega+i0^{+}$. In order to ease geometrical
complexity, we refrain from confining the complex fermions to only a reduced
``positive'' region of momentum space. This comes at the expense of
additional anomalous anticommutators like, e.g., $\{d_{1{\bf
k}}^{\phantom{\dagger}},d_{2{\bf k}'}^{\phantom{\dagger}}\}=\delta_{-{\bf
k},{\bf k}'}$ and their corresponding contractions. Simple algebra yields
\begin{align}
\chi({\bf q},z)
& =\chi^{\mathrm{ph}}({\bf q},z)+
\chi^{\mathrm{pp}}({\bf q},z)\label{eq:6}\\
\chi^{\mathrm{ph}}({\bf q},z)
& =\frac{1}{N}\sum_{{\bf k}}(\epsilon_{{\bf k}+{\bf q}}{+}
\epsilon_{{\bf k}})^{2}
\frac{f_{{\bf k}+{\bf q}}(T){-}f_{{\bf k}}(T)}
{z-\epsilon_{{\bf k}+{\bf q}}+\epsilon_{{\bf k}}}\nonumber \\
\chi^{\mathrm{pp}}({\bf q},z)
& =\frac{1}{2N}\sum_{{\bf k}}(\epsilon_{{\bf k}+{\bf q}}{-}
\epsilon_{{\bf k}})^{2}
\left\{ [f_{{\bf k}+{\bf q}}(T){+}f_{{\bf k}}(T){-}1]\right.\nonumber \\
& \phantom{aaa}\times\left.\left(
\frac{1}{z-\epsilon_{{\bf k}+{\bf q}}-\epsilon_{{\bf k}}}{-}
\frac{1}{z+\epsilon_{{\bf k}+{\bf q}}+\epsilon_{{\bf k}}}\right)\right\} \,,
\nonumber 
\end{align}
where the superscripts ph(pp) indicate particle-hole (particle-particle) or,
synonymous intra(inter)band type of intermediate states of the fermions,
$f_{{\bf k}}(T)=1/(e^{\epsilon_{{\bf k}}/T}+1)$ is the Fermi function. This
concludes the formal details for $T\lesssim T^{\star}$.

\subsection{Random gauge for $T\gtrsim T^{\star}$\label{subsec:rand}}

In a random gauge configuration, translational invariance of the Majorana
system is lost, and we resort to a numerical approach in real space. This has
been detailed extensively for 1D \cite{Metavitsiadis2017a,
Metavitsiadis2020b} and 2D \cite{Pidatella2019, Metavitsiadis2017,
Metavitsiadis2020a} models and is only briefly reiterated here for
completeness sake. First a spinor $A_{ \text{ \ensuremath{ \sigma}}}^{
\dagger} =( a_{1}\dots a_{{\bf l}}\dots a_{N},c_{1}\dots c_{{\bf l}+{\bf
r}_{x}}\dots c_{N})$, comprising the Majoranas on the $2N$ sites of the
lattice is defined.  Using this, the energy density $h_{{\bf q}}$ and the
Hamiltonian (\ref{eq:2}), i.e., $h_{{\bf 0}}$, are rewritten as $h_{{\bf
q}}={\bf A}^{\dagger}{\bf g}_{{\bf q}}{\bf A}/2$.  Bold faced symbols refer
to vectors and matrices, i.e., ${\bf g}_{{\bf q}}$ is a $2N\times2N$
array. Next a spinor $D_{\sigma}^{\dagger} = (d_{1}^{\dagger}\dots
d_{N}^{\dagger}, d_{1}^{\phantom{\dagger}} \dots d_{N}^{\phantom{\dagger}})$
of $2N$ complex fermions is defined by ${\bf D}={\bf F}{\bf A}$ using the
unitary (Fourier) transform ${\bf F}$. The latter is built from two disjoint
$N\times N$ blocks $I_{\sigma\rho}^{i=1,2} = e^{-i{\bf k}_{\sigma}\cdot{\bf
R}_{\rho}^{i}} / \sqrt{N}$, with ${\bf R}_{\rho}^{i}={\bf l}$ and ${\bf
l}+{\bf r}_{x}$, for $a$- and $c$-Majorana lattice sites, respectively. ${\bf
k}$ is chosen such, that for each ${\bf k}$, there exists one $-{\bf k}$,
with ${\bf k}\neq-{\bf k}$. Finally, for convenience, ${\bf F}$ is rearranged
such as to associate the $d_{1}^{\dagger}\dots d_{N}^{\dagger}$ with the
$2\,(N/2)=N$ ``positive'' ${\bf k}$-vectors. With this
\begin{equation}
h_{{\bf q}}={\bf D}^{\dagger}{\bf \tilde{g}}_{{\bf q}}\,{\bf D}/2\,,
\label{eq:7}
\end{equation}
where $\tilde{{\bf o}}={\bf F}{\bf o}{\bf F}^{\dagger}$. We emphasize, that
(i) ${\bf F}$ does \emph{not} diagonalize $h_{{\bf q}}$ and (ii) that in
general, the $2N\times2N$ matrices of Fourier transformed operators
$\tilde{{\bf o}}$ will contain particle number non-conserving entries of
${\bf D}$ fermions.

As for the case of the homogeneous gauge in Sec. \ref{subsec:homg}, the
energy density susceptibility $\chi({\bf q},z)$ for a \emph{particular} gauge
sector $\{\eta_{{\bf l}}\}$ is obtained by analytic continuation from the
imaginary time density Green's function
\begin{align}
\chi({\bf q},\tau)
& =\langle T_{\tau}(h_{{\bf q}}(\tau)
h_{-{\bf q}})\rangle_{\{\eta_{{\bf l}}\}}\nonumber \\
& =\frac{1}{4}\langle T_{\tau}[({\bf D}^{\dagger}
{\bf \tilde{g}}_{{\bf q}}{\bf D})(\tau)
({\bf D}^{\dagger}{\bf \tilde{g}}_{{\bf q}}
{\bf D})^{\dagger}]\rangle_{\{\eta_{{\bf l}}\}}\,.
\label{eq:8}
\end{align}
This is evaluated using Wick's theorem for quasiparticles ${\bf T}={\bf
U}{\bf D}$, referring to a $2N\times2N$ Bogoliubov transformation ${\bf U}$,
determined numerically for a given distribution $\{\eta_{{\bf l}}\}$, such as
to diagonalize ${\bf \tilde{g}}_{{\bf 0}}$, i.e., $({\bf U}{\bf
\tilde{g}}_{{\bf 0}} {\bf U}^{\dagger} )_{ \rho\sigma} = \delta_{ \rho\sigma}
\epsilon_{\rho}$, with $\epsilon_{ \rho} =( \epsilon_{1} \dots \epsilon_{N},
-\epsilon_{1} \dots -\epsilon_{N})$. We get
\begin{align}
\chi({\bf q},z) & =\sum_{\rho\sigma}\Pi_{\sigma\rho}(z)
w_{\sigma\rho,{\bf q}}[w_{\bar{\rho}\bar{\sigma},
{\bf q}}^{\star}-w_{\sigma\rho,{\bf q}}^{\star}]\,,\nonumber \\
\Pi_{\sigma\rho}(z) &
=\frac{f_{\sigma}(T)-f_{\rho}(T)}{z-\epsilon_{\sigma}+
\epsilon_{\rho}}\,,
\label{eq:9}\\
w_{\rho\sigma,{\bf q}} &
=(\frac{1}{2}{\bf U}\tilde{{\bf g}}_{{\bf q}}
{\bf U}^{\dagger})_{\rho\sigma}\,,\nonumber 
\end{align}
where $f_{\sigma}(T)=1/(e^{\epsilon_{\sigma}/T}+1)$, and overbars refer to
swapping the upper and lower half of the range of 2$N$ indices,
e.g., $\bar{\rho} = \rho \mp N$ for $\rho\gtrless N$.

As a final step, $\chi({\bf q},z)$ from Eq. (\ref{eq:9}) is averaged over a
sufficiently large number of random distributions $\{\eta_{{\bf l}}\}$.  This
concludes the formal details of the evaluation of the energy density
susceptibility for $T\gtrsim T^{\star}$.

\subsection{Diffusion Kernel\label{subsec:Dqw}}

We will relate $\chi({\bf q},z)$ to a diffusion kernel $D({\bf q},z)$ by the
following phenomenological \emph{Ansatz}, rooted in hydrodynamic theory and
memory function approaches \cite{Forster1975}
\begin{equation}
\chi({\bf q},z)=\chi_{{\bf q}}\,\frac{iq^{2}D({\bf q},z)}
{z+iq^{2}D({\bf q},z)}\,.
\label{eq:10}
\end{equation}
This should be viewed as a definition of $D({\bf q},z)$ and a static
energy-density susceptibility $\chi_{{\bf q}}$. Neither does this take into
account fine details concerning differences between static, adiabatic, or
isolated susceptibilities nor does it formulate the momentum scaling in terms
of harmonics of the honeycomb lattice instead of the simpler factor of
$q^{2}$. The latter implies, that the momentum dependence of $D({\bf q},z)$
is adapted best to the small $q$ regime.

Since by construction of (\ref{eq:10}), $D({\bf q},z)$ has a proper spectral
representation, $\chi_{{\bf q}}$ results from the sum rule
\begin{equation}
\chi_{{\bf q}} = \int_{-\infty}^{\infty}\frac{d\omega}
{\pi\omega}\mathrm{Im}[\chi({\bf q},\omega_{+})]\,,
\label{eq:11}
\end{equation}
with $\chi({\bf q},\omega_{+})=\chi({\bf q},\omega+i0^{+})$. Introducing a
normalized susceptibility $\bar{\chi}({\bf q},\omega_{+})=\chi({\bf
q},\omega_{+})/\chi_{{\bf q}}$, we will extract the diffusion kernel from
\begin{equation}
D({\bf q},\omega_{+}) = \frac{1}{iq^{2}}\frac{\omega\,
\bar{\chi}({\bf q},\omega_{+})}{1-\bar{\chi}({\bf q},\omega_{+})}\,.
\label{eq:12}
\end{equation}

\begin{figure}[tb]
\centering{}\includegraphics[width=0.7\columnwidth]{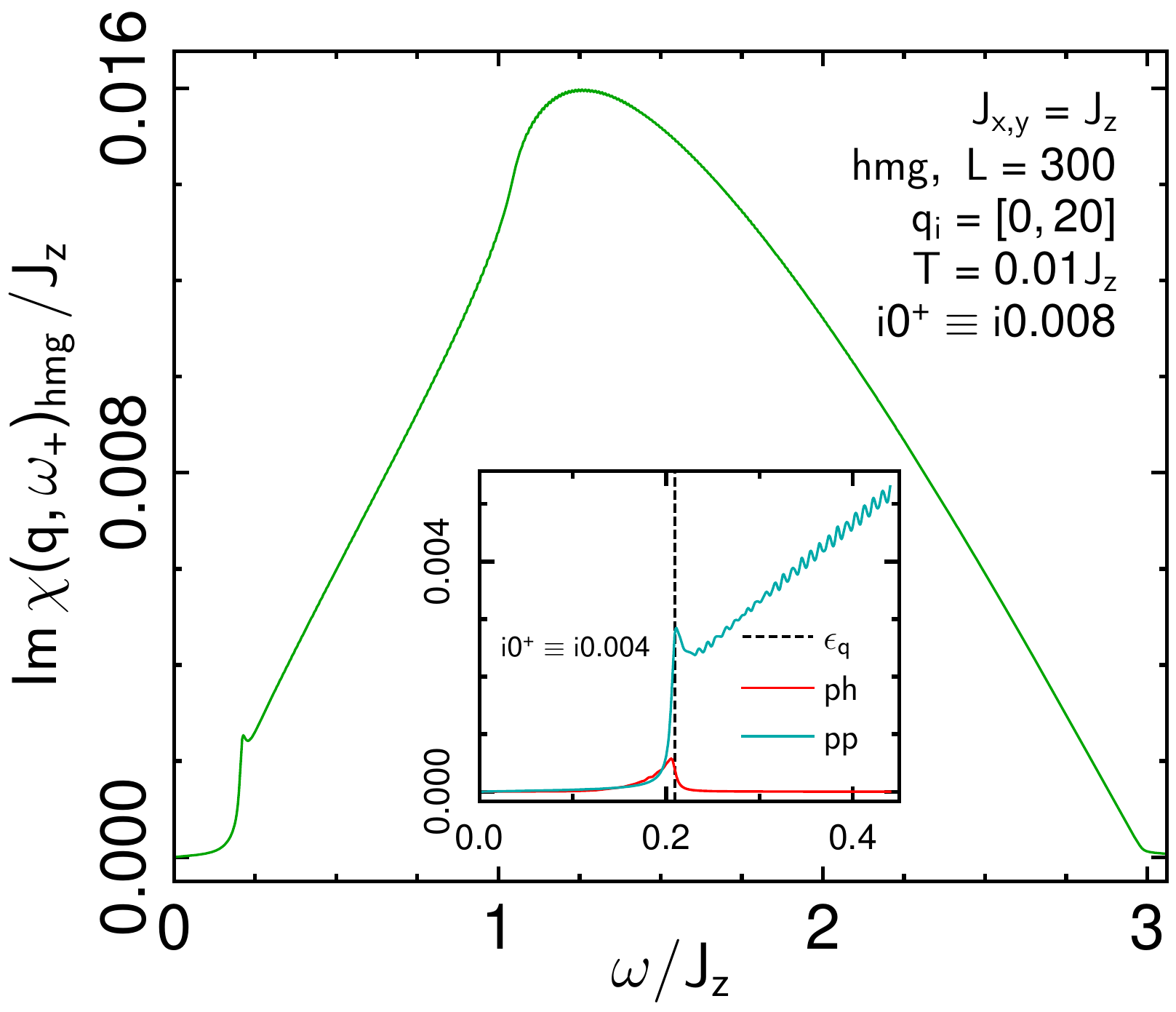}
\caption{\label{fig:1}Spectrum, $\mathrm{Im}\chi({\bf q},\omega_{+})$, of
energy density susceptibility versus $\omega$ at fixed small ${\bf q}$, for
low $T=0.01J_{z}\lesssim T^{\star}$, using Eq. (\ref{eq:6}) for homogeneous
ground state gauge. System size $L^2 = 300\times 300$. 
Momentum ${\bf q}=2\pi/L\sum_{j=1,2}q_{ij}{\bf G}_{j}$.
Inset: Blow up of low-$\omega$ region with reduced imaginary broadening
$i0^{+}$. Dashed black: upper ph-continuum bound at $T=0$.}
\end{figure}

\section{Results\label{sec:Results}}

We now discuss the density dynamics as obtained from the previous
sections. First we consider the low-$T$ behavior, $T\lesssim T^{\star}$,
using the homogeneous gauge ground state. As from Eq.~(\ref{eq:6}), 
$\chi({\bf q},z)$ sums two channels: (i) particle-hole (ph) and (ii)
particle-particle (pp) excitations. Their spectral support is
$0{<}{|}\omega{|}{<}\epsilon_{\tilde{{\bf q}}}$ for (ph) and
$\epsilon_{\tilde{{\bf q}}}{<}{|}\omega{|}{\lesssim}\max(2\epsilon_{{\bf
k}}){=}3J$ at ${|}{\bf q}{|}{\ll}1$ for (pp), where $\tilde{{\bf q}}{=}{\bf
q}{+}{\bf k}_{D}$ refers to the wave vector with respect to the location of
the Dirac cone. A typical spectrum $\mathrm{Im}\chi({\bf q},\omega_{+})$ is
shown in Fig. \ref{fig:1} at small, albeit finite ${\bf q}$. The inset
depicts a blowup of the low-$\omega$ region dissecting the spectrum into its
ph and pp contributions. While the imaginary broadening in the inset is
already reduced such that finite size oscillations start to show, the
pp-channel still exhibits some weight below its cut-off at
$\epsilon_{\tilde{{\bf q}}}\simeq0.208J_{z}$. This will vanish as
$\mathrm{Im}z\rightarrow0$. For the ph-channel however, the spectral weight
in this energy range is not a finite size effect.  Due to the Dirac cone, the
Fermi volume shrinks to zero in the Kitaev model as $T{\rightarrow}0$, i.e.,
occupied states only stem from a small patch with $\epsilon_{{\bf k}}\lesssim
T$ around the Dirac cone. Therefore, the weight of the ph-channel decreases
rapidly to zero as $T{\rightarrow}0$. In this regime and for small ${\bf q}$,
because of the linear fermion dispersion close to the cones, only a narrow
strip of order $\omega\in[\max(0,\epsilon_{\tilde{{\bf q}}} -2T ),
\epsilon_{\tilde{{\bf q}}}]$ from the spectral support dominates the
ph-continuum. At the upper edge of its support the ph DOS is singular. The
inset of Fig. \ref{fig:1} is consistent with this, considering the finite
system size and imaginary broadening used.

\begin{figure}[tb]
\begin{centering}
\includegraphics[width=0.8\columnwidth]{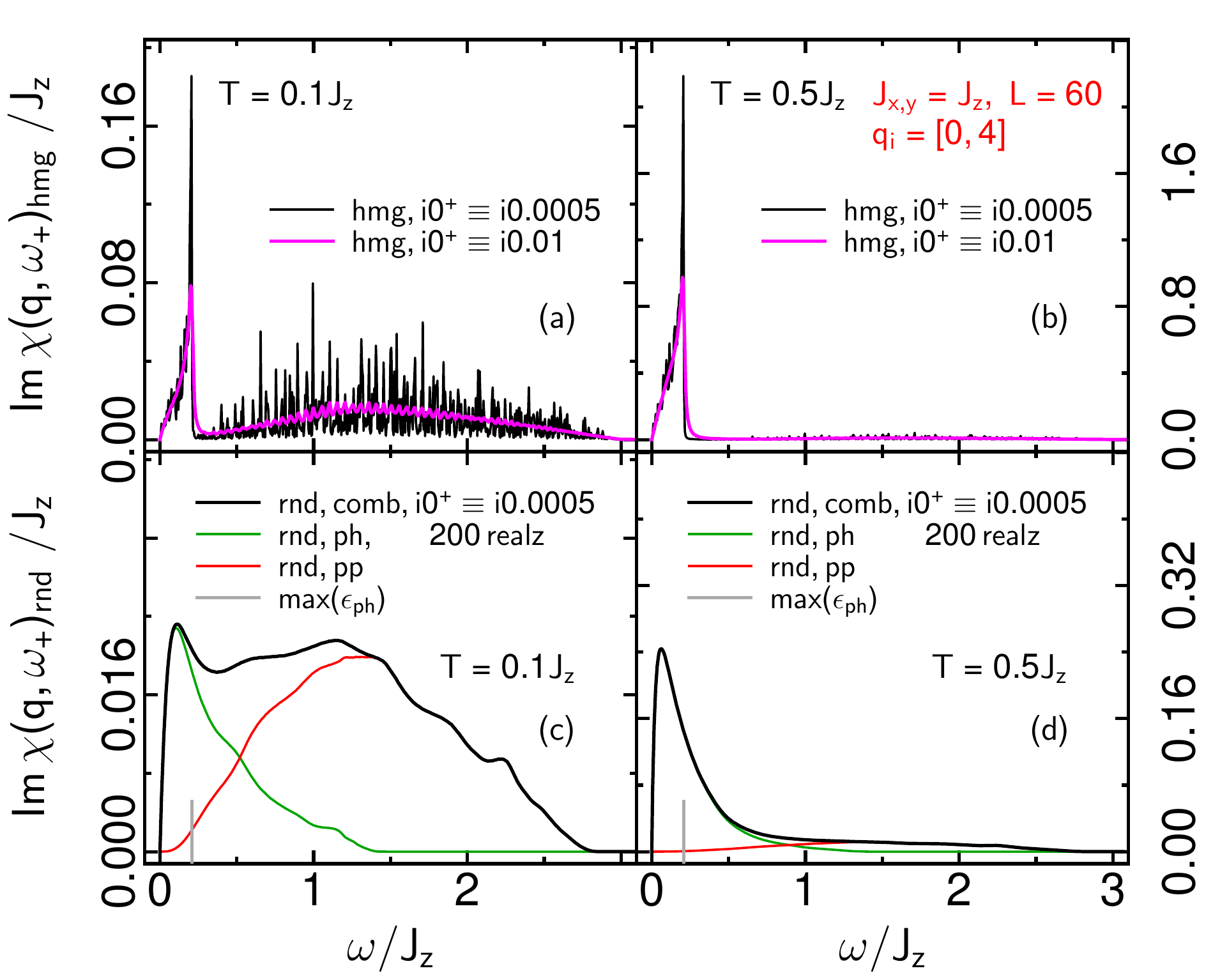} 
\end{centering}
\caption{\label{fig:2}Spectrum, $\mathrm{Im}\chi({\bf q},\omega_{+})$, of
energy density susceptibility versus $\omega$ at fixed small ${\bf q}= 2 \pi
/ L \sum_{j=1,2} q_{ij} {\bf G}_{j}$, contrasting homogeneous (a,b) versus
random (c,d) gauge on all identical system sizes $L^{2}=60\times60$ for two
identical temperatures $T=0.1J_{z}$ (a,c) and $T=0.5J_{z}$ (b,d). Solid
black, green, and red lines in panels (c,d): total, ph-, and
pp-spectrum. Extended vertical gray ticks in panels (c,d): upper ph-continuum
bound at $\boldsymbol{q}$ in homogeneous gauge. Imaginary smoothing $i0^{+}$
for homogeneous case, both, identical (solid black) and significantly larger
than random gauge case (solid magenta). Note the different y-axis scales.}
\end{figure}

Regarding the pp-channel, the complete two-particle continuum is unoccupied
and available for excited states as $T{\rightarrow}0$. This leads to the
broad spectral hump seen in Fig. \ref{fig:1}, which extends out to $\max ( 2
\epsilon_{{\bf k}}) = 3 J_{z}$, at $J_{x,y}=J_{z}$ and is two orders of
magnitude larger than the ph-process at this temperature.

A fingerprint of potentially diffusive relaxation of density modes at finite
momentum ${\bf q}$ is the near-linear behavior of $\mathrm{ Im } \chi( {\bf
q},\omega_+) \sim \chi_{\bf q} \,\omega / (D q^2)$ at small
$\omega$. Definitely, this should neither be expected, nor is it observed in
Fig. \ref{fig:1}, since for $T \lesssim T^\star$, the density dynamics is set
by coherent two-particle excitations of the Dirac fermions in the
homogeneous gauge state.

For the remainder of this work, we therefore now turn to temperatures above
the flux proliferation, i.e., $T\gtrsim T^{\star}$, using  random gauge
states.  To begin, and in Fig. \ref{fig:2}, we first describe the impact of the
random gauges, by contrasting the dynamic density susceptibilities against
each other with, and without random gauges, for otherwise identical system
parameters, and for two different temperatures, $T=0.1$ and 0.5, in
Figs. \ref{fig:2}(a,c) and (b,d), respectively. Note, that while the linear
system size is smaller by a factor of five with respect to Fig. \ref{fig:1},
the wave vector has also been rescaled accordingly.  Therefore these two
figures can be compared directly. Obviously, in the homogeneous gauge,
significant degeneracies, even on 60$\times$60 lattices, lead to visible
discretization effects. Therefore, in Figs.  \ref{fig:2}(a,b) we include
spectra with an imaginary broadening, increased relative to
Figs. \ref{fig:2}(c,d), for a better comparison with the latter.

\setcounter{figure}{3} 
\begin{figure*}[t]
\begin{centering}
\includegraphics[width=2\columnwidth]{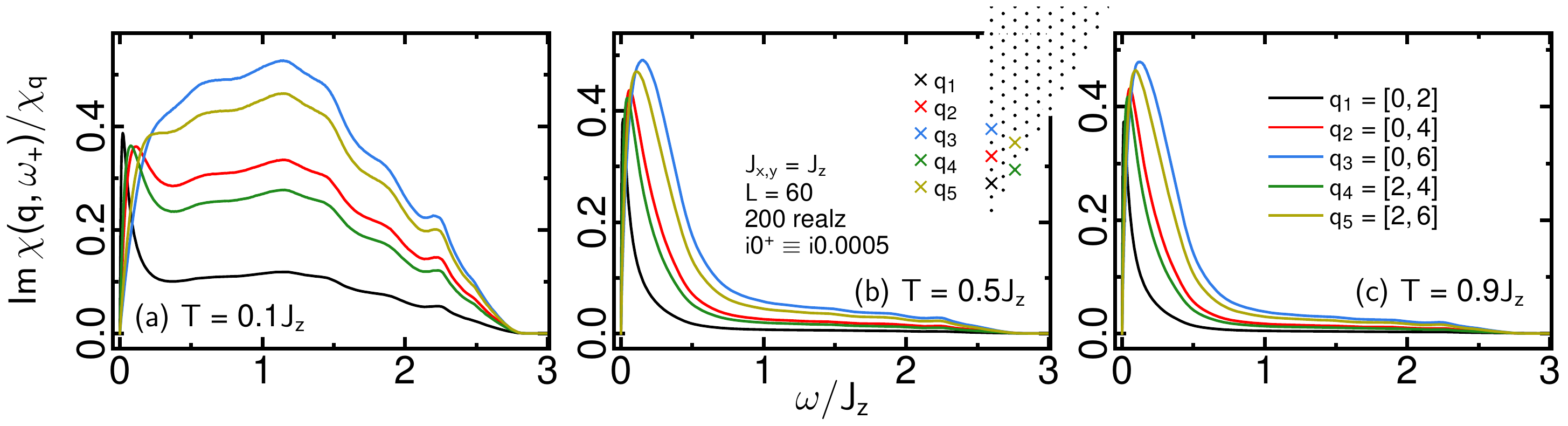} 
\end{centering}
\caption{\label{fig:4}Spectrum of the normalized density susceptibility
$\mathrm{Im}\chi({\bf q},\omega_{+})/\chi_{\boldsymbol{q}}$ in the random
gauge state, with 200 realizations, versus $\omega$ for various temperatures
$T/J_{z}=0.1,$ 0.5, and 0.9 in panels (a), (b), and (c), for various small
momenta ${\bf q}=2\pi/L\sum_{j=1,2}q_{ij}{\bf G}_{j}$ within the small-$q$
region, inset panel (b), of the irreducible wedge of the BZ for
$L^{2}=60\times60$ sites and for imaginary broadening $i0^{+}$.}
\end{figure*}

\setcounter{figure}{2} 
\begin{figure}[b]
\begin{centering}
\includegraphics[width=0.75\columnwidth]{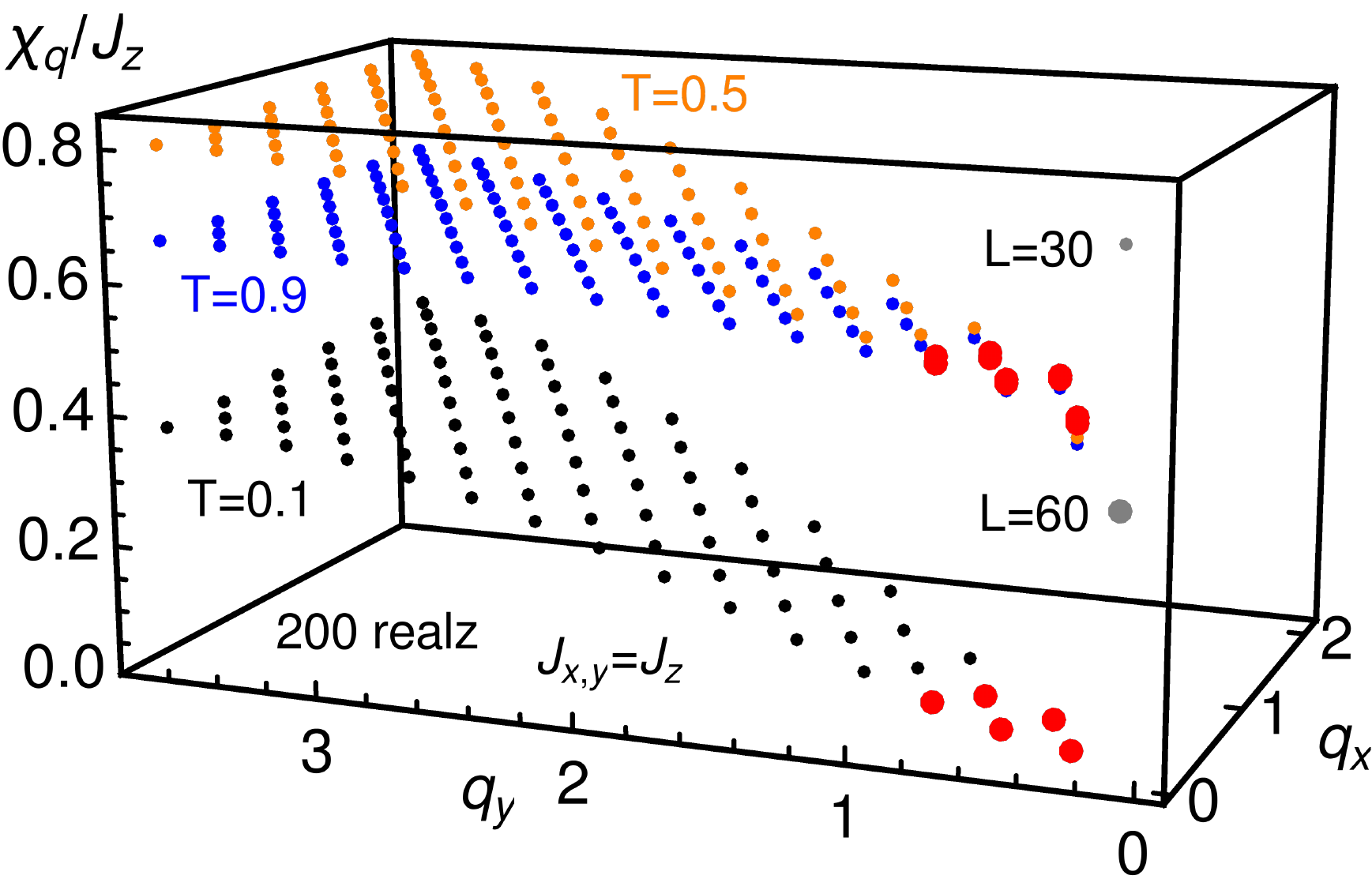} 
\end{centering}
\centering{}
\caption{\label{fig:3}Static susceptibility $\chi_{\boldsymbol{q}}$ versus
momentum in the random gauge state, with 200 realizations. Small solid dots:
all of irreducible wedge of the BZ at $L^{2}=30\times30$ for three
temperatures $T/J_{z}=0.1$, 0.5, and 0.9 (black, orange, and blue). Big solid
red dots, finite size effects: $\chi_{\boldsymbol{q}}$ at low$-q$ for
$L^{2}=60\times60$ and simultaneously $\boldsymbol{q} \in 30 \times 30$ BZ.}
\end{figure}

Several features can be observed. First, while the ph-channel in the uniform
gauge clearly displays the singular behavior at $\omega=\epsilon_{\tilde{{\bf
q}}}$, mentioned earlier and visible because of the elevated temperatures, in
the random gauge case, it displays a smooth peak. Second, as can be read of
from the $y$-axis, the weight of the ph-channel strongly increases with
increasing $T$. For the temperatures depicted, the pp-channel is much less
$T$-dependent. Third, the ph-, versus the pp-contributions to
$\mathrm{{Im}}\chi({\bf q},\omega_+)$ cannot only be dissected in the uniform
gauge by virtue of Eqs. (\ref{eq:6}), but also for the random gauge,
Eq. (\ref{eq:9}) can be decomposed into addends with
$\epsilon_{\sigma}\epsilon_{\rho}\gtrless0$. This evidences, that in the
latter case, the ph-spectrum changes completely.  As is obvious from
Figs. \ref{fig:2}(c,d), the ph-channel spreads into a broad feature,
extending over roughly the entire one-particle energy range. The pp-channel
on the other hand seems less affected by the gauge disorder, with a shape
qualitatively similar to that in the gauge ground state, as can be read off
by comparing Figs. \ref{fig:2}(a,c).

Most remarkably, for intermediate temperatures, as in Fig. \ref{fig:2}(d) at
$T/J_{z}=0.5$, the overall shape of the spectrum is very reminiscent of a
diffusion-pole behavior at fixed momentum, i.e., $\mathrm{Im} \chi( {\bf q},
\omega) \propto \omega \Gamma/(\omega^2+\Gamma^2)$, with some relaxation rate
$\Gamma$. To clarify this in more detail, we therefore proceed and analyze
$\chi({\bf q},z)$ in terms of the hydrodynamic expression Eq. (\ref{eq:10}).

To this end, we first extract the static susceptibility
$\chi_{{\bf q}}$, performing the sum rule of Eq. (\ref{eq:11}) via numerical
integration, using $\mathrm{{Im}}\chi({\bf q},z)$ from the corresponding
random gauge states. The results are shown in Fig. \ref{fig:3}, spanning an
irreducible $\boldsymbol{q}$-wedge of the BZ. Since energy conservation
renders the dynamic density response singular at $\boldsymbol{q} =
\boldsymbol{0}$, this momentum will be excluded hereafter. Obviously
$\chi_{{\bf q}}$ is a smooth and featureless function. The figure also
contrasts 30$\times$30 against 60$\times$60 systems at selected momenta. The
finite size effects are small.

Next, and in Fig. \ref{fig:4}, we consider the global variation with
momentum, of the {\em normalized} spectrum of the dynamical energy density
susceptibility versus $\omega$. Since our focus is on the hydrodynamic
regime, we remain with small momenta. These momenta are indicated on a
fraction of an irreducible wedge of the BZ in Fig. \ref{fig:4}(b). A spacing
of $\mathrm{Mod}(2)$ of the momenta has been chosen to allow for later
analysis of finite size effects in comparison to systems with a linear
dimension smaller by a factor of 2. The spectra show significant changes with
momentum. First, the low-$\omega$ contributions, which stem primarily from
the ph-channel show a broadening of their support with increasing
$|\boldsymbol{q}|$. Second, the spectral range of pp-excitations displays a
global increase of weight with $|\boldsymbol{q}|$. Finally, at intermediate
and elevated $T$ in Fig. \ref{fig:4}(b) and (c), the former effect dominates
the latter regarding the global shape of the spectrum.

\setcounter{figure}{4} 
\begin{figure*}[t]
\centering{}\includegraphics[width=2\columnwidth]{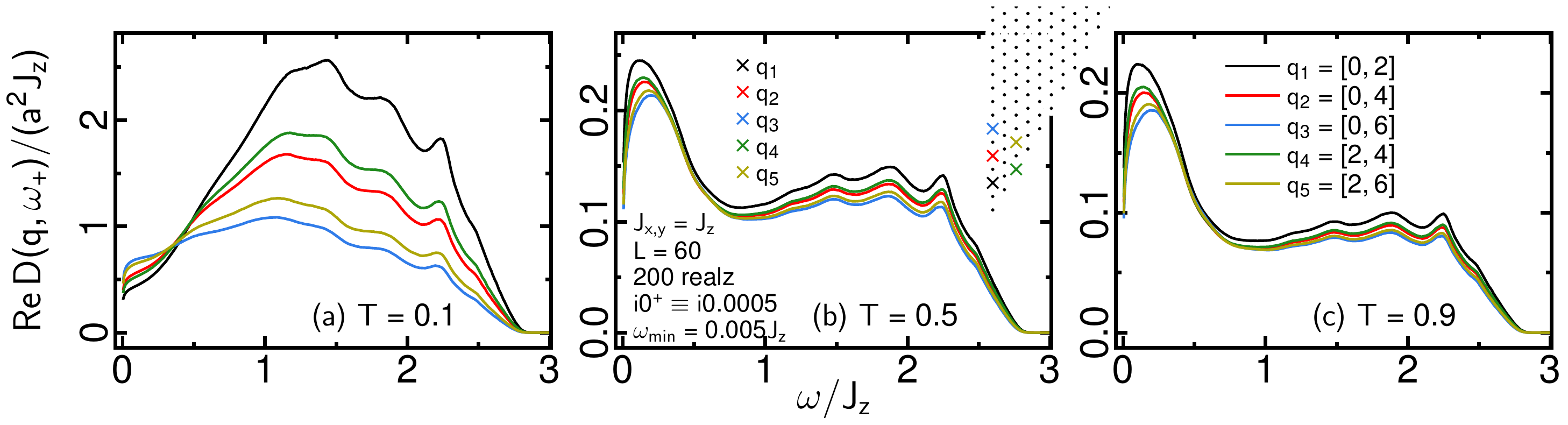}
\caption{\label{fig:5}Real part of the diffusion kernel $\mathrm{Re}D({\bf
q},\omega_{+})$ in the random gauge state, with 200 realizations, versus
$\omega\ge.005J_{z}$ for various temperatures $T/J_{z}=0.1,$ 0.5, and 0.9 in
panels (a), (b), and (c), for various small momenta ${\bf q} = 2 \pi / L
\sum_{j=1,2} q_{ij} {\bf G}_{j}$ within the small-$q$ region, inset panel
(b), of the irreducible wedge of the BZ for $L^{2}=60\times60$ sites and for
imaginary broadening $i0^{+}$.}
\end{figure*}

\begin{figure}[b]
\begin{centering}
\includegraphics[width=0.5\columnwidth]{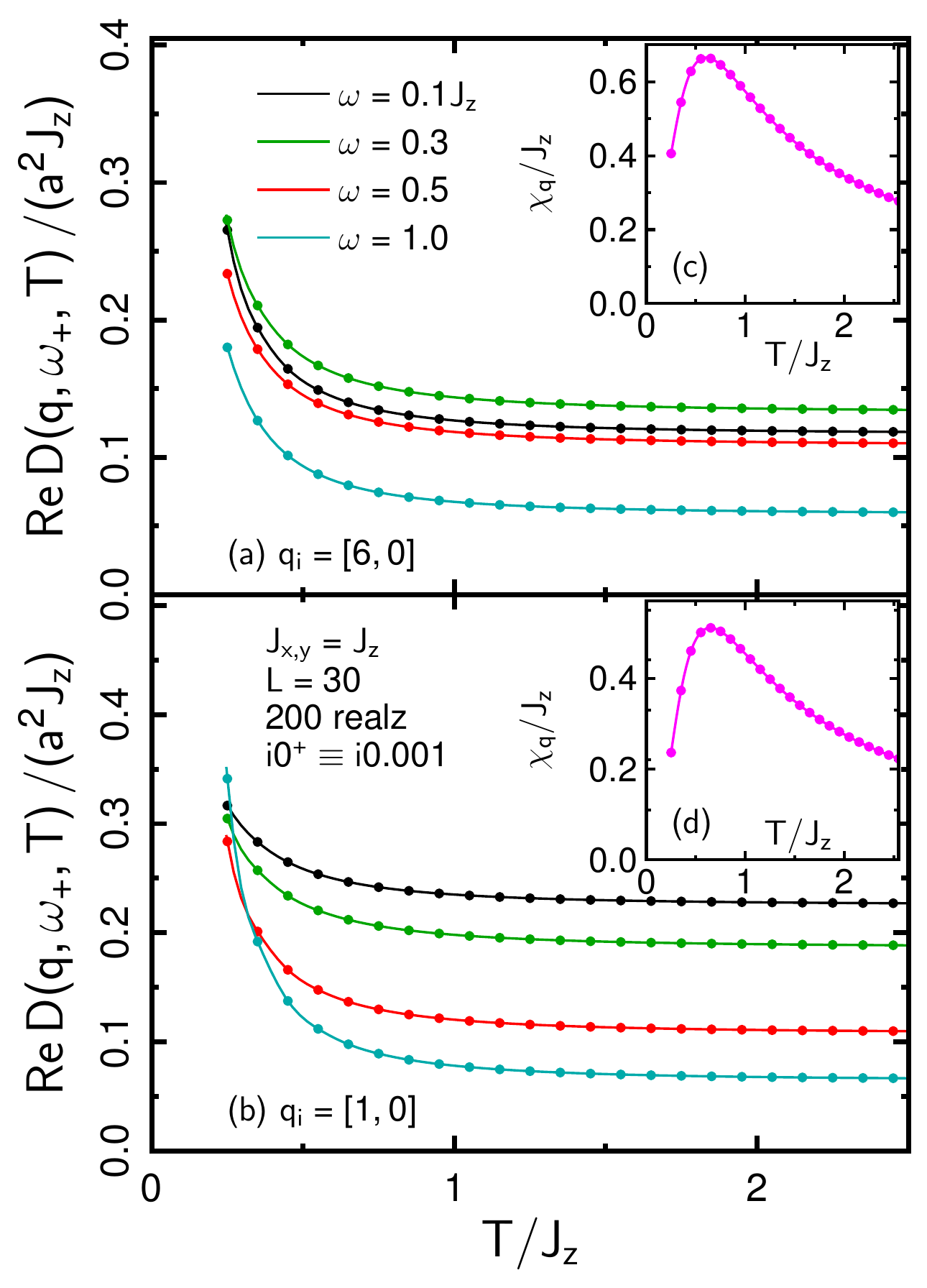} 
\end{centering}
\caption{\label{fig:6}Real part of the diffusion kernel $\mathrm{Re}D({\bf
q},\omega_{+},T)$ in the random gauge state, with 200 realizations, versus
$T/J_{z}$ at ${\bf q}=2\pi/L\sum_{j=1,2}q_{ij}{\bf G}_{j}$ at (a)
$\boldsymbol{q}_{i} = [6,0]$ and (b) $[1,0]$, and various fixed energies
$\omega$, for $L^{2}=30\times30$ sites and imaginary broadening
$i0^{+}$. Insets (c,d): static susceptibility $\chi_{\boldsymbol{q}}(T)$
versus $T/J_{z}$ at identical $L$ and $\boldsymbol{q}$.}
\end{figure}

Now we extract the diffusion kernel $D({\bf q},\omega_{+})$ as in
Eq. (\ref{eq:12}). Its real part is depicted in  Fig. \ref{fig:5} versus
$\omega$ for identical momenta and temperatures as in Fig.
\ref{fig:4}. Clearly, at low $T$, Fig. \ref{fig:5}(a), the diffusion kernel
displays significant $\boldsymbol{q}$-dependence, implying that the energy
currents are not proportional to the gradient $\nabla_{ \boldsymbol{l}} h_{
\boldsymbol{l}}$ of the energy density and therefore a simply hydrodynamic
picture is not applicable. In contrast to that, at intermediate and high $T$,
Fig. \ref{fig:5}(b) and (c), the diffusion kernel is approximately momentum
independent $\mathrm{Re}D(|{\bf q}| \ll 1, \omega_{+}) \simeq \mathrm{Re} D(
\omega_{+})$.  This is consistent with Fick's law regarding $\boldsymbol{
q}$-scaling.  However, the diffusion process remains retarded, although not
very strong, displaying a shoulder in the pp-range and a peak at low $\omega$
in the ph-range.

\begin{figure}[tb]
\begin{centering}
\includegraphics[width=0.7\columnwidth]{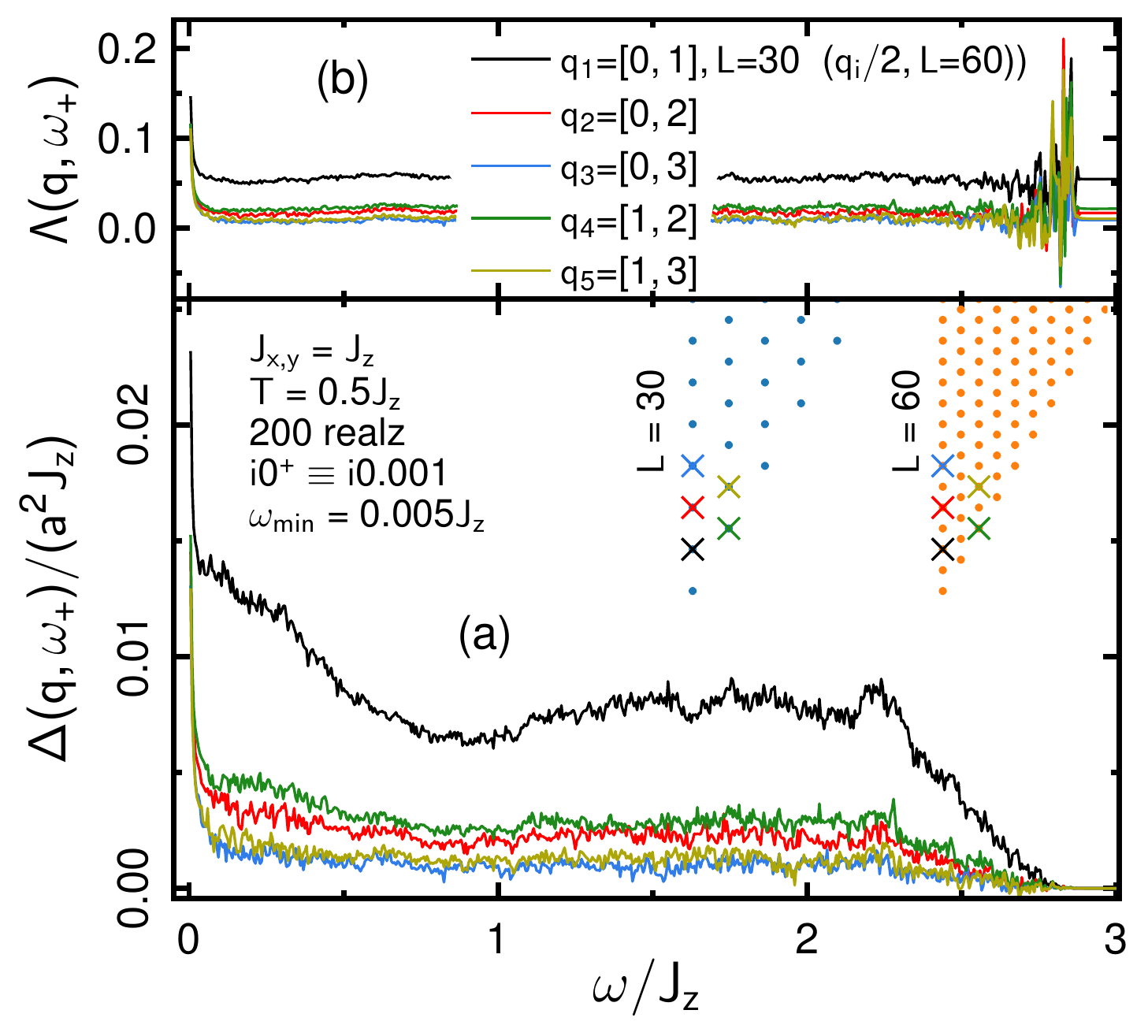} 
\end{centering}
\caption{\label{fig:7}(a) Absolute and (b) relative difference of real part
of the diffusion kernel $\mathrm{Re}D({\bf q},\omega_{+})$ in the random
gauge state, with 200 realizations, versus $\omega\ge0.005J_{z}$ at
$T/J_{z}=0.5$, for various small momenta ${\bf q} = 2 \pi / L \sum_{ j=1,2}
q_{ij} { \bf G}_{j}$ on $L^{2}=30\times30$ and $\boldsymbol{ q}_{ i}
\rightarrow 2 \boldsymbol{ q}_{i}$ on $L^{2}=60\times60$, within the
small-$q$ region of insets panel (a), of the irreducible wedge of the BZ and
for imaginary broadening $i0^{+}$.
%\vspace*{-.4cm}
}
\end{figure}

As $\omega/J_{z}\rightarrow0$, the numerical accuracy of the transform
Eq. (\ref{eq:12}), comprising small numbers in the numerator and denominator,
is an issue with respect to system size and imaginary broadening and we have
to remain with $\omega \geq \omega_{ \mathrm{ min}} = 0.005 J_{z}$ for the
parameters used.  See also the discussion of Fig. \ref{fig:7}. In view of the
steep slope in this regime, it may be that on any \emph{finite} system
$\mathrm{Re} D(|{\bf q}|\ll1,\omega=0)$ vanishes singularly below the
unphysically small energy scale $\omega_{\mathrm{min}}$, while in the
thermodynamic limit $\mathrm{Re}D( | {\bf q} | \ll 1, \omega = 0 )$ is of the
order of the low-energy peak height. This behavior is very reminiscent of
similar findings for the thermal conductivity, see
Ref.~\cite{Metavitsiadis2017} and App.~\ref{app1}.  For $\omega / J_{z}
\gtrsim 2.8$, the spectral support terminates and the diffusion kernel turns
purely imaginary $\propto \omega^{ -1 }$.  In conclusion, at not too low
temperatures and not too short time scales, gauge disorder in the Kitaev
magnet leads to an energy density dynamics, very similar to conventional
diffusion, regarding its momentum scaling, with, however, some retardation
remaining. This should be contrasted with the underlying spin model being
a translationally invariant system.

Turning to the temperature dependence, we consider two representative momenta
$\boldsymbol{q}$ and several energies. The corresponding diffusion kernel
$\mathrm{ Re} D( {\bf q}, \omega_{+})$ and the static susceptibility $\chi_{
\boldsymbol{q}}$ are shown versus $T$ in Figs.  \ref{fig:6}(a), (b) and (c),
(d), respectively. The temperature range has been truncated deliberately to $T / J_{z}
\gtrsim 0.25$, since below such temperatures, and from the ${\bf q}$-scaling
in Fig. \ref{fig:5} the density dynamics is far from hydrodynamic. The figure
clearly demonstrates, that for $T/J_{z} \gtrsim 0.5$, the diffusion kernel
rapidly settles onto some almost constant value, set by energy and
momentum. This is consistent with Fig. \ref{fig:5}, which displays only weak
overall change between the diffusion kernels for the two temperatures of
panels (b) and (c). As a consequence, the global $T$-dependence of $\chi(
\boldsymbol{q}, z)$ is essentially set by the static energy susceptibility.
As the insets show, the latter exhibits a maximum in the vicinity of $T /
J_{z} \sim 1$. For $T / J_{z} \gg 1$, $\chi_{{\bf q}}$ approaches its
classical limit, decaying $\propto T^{-1}$, which can also be read of from
Eqs. (\ref{eq:11},\ref{eq:9}). Such behavior is typical also for other static
susceptibilities of spin systems.

In closing, we provide some measure of the finite size effects involved in
our calculations. To this end we consider both, the absolute and the relative
difference between the diffusion kernels, $\Delta({\bf q}, \omega ) = |
\mathrm{Re}[D_{30{\times}30}({\bf q}, \omega_{+} ) - D_{60{\times}60}({\bf
q}, \omega_{+})]|$ and $\Lambda({\bf q}, \omega) = 2 \Delta({\bf q}, \omega )
/ | \mathrm{Re}[D_{30{\times}30}({\bf q}, \omega_{+})+D_{60{\times}60}({\bf
q},\omega_{+})]|$, respectively, on $N=30{\times}30$ and $60{\times}60$ site
systems, for identical wave vectors. Regarding the latter, this implies a
factor of $2$ difference between their integer representation in terms of
${\bf G}_{1,2}$. The differences are shown in Fig. \ref{fig:7}.  They display
statistical noise from the finite number of random gauge realizations and
remain acceptably small for all $\omega$. Only for $0\approx\omega\ll0.01$,
where $\Lambda({\bf q}, \omega)$ is of $O(10\%)$, the error is not of finite
size, or statistical origin, but rather stems from the systematic numerical
inaccuracies, mentioned in the preceding, of the denominator in
Eq. (\ref{eq:12}) with $\chi({\bf q},\omega_{+})$ obtained from
Eq. (\ref{eq:9}) as $\omega\rightarrow0$.  In turn, $D({\bf q},\omega_{+})$
at very low $0\approx\omega\ll0.01$ may be inacurate by $\sim10\%$. As to be
expected, the actual finite size errors are largest for the smallest wave
vector.

\section{Summary\label{sec:Summary}}

In summary, above an intermediate temperature scale $T \sim 0.5J$,
which is still well below the classical limit, the energy density in
Kitaev magnets at finite momentum relaxes remarkably similar to
diffusion in random media, with, however, a clearly notable
difference. Namely, while the momentum scaling is practically
hydrodynamic $\propto q^2$, the diffusion kernel is not completely
energy independent, i.e., it displays some retardation within its
support. The origin of the latter can be traced back to the presence
of two distinct relaxation channels for the energy density, comprising
particle-hole and particle-particle excitations of the Dirac
fermions. Both propagate in a strongly disordered landscape, created
by thermally induced gauge excitations. Their combined effect, however,
does not lead to a constant diffusion rate. At extremely low energies
we observe a dip in the diffusion kernel. This is consistent with
similar claims for the dynamical thermal conductivity, to which we
find that our results connect consistently via generalized Einstein
relations.  Future analysis, focusing on the real space, instead of
the momentum dependence of energy-density relaxation could be
interesting in order to predict finite temperature quench dynamics.

\emph{Acknowledgments}: This work has been supported in part by the DFG
through project A02 of SFB 1143 (project-id 247310070), by Nds. QUANOMET
(project NP-2), and by the National Science Foundation under Grant No. NSF
PHY-1748958. W.B. also acknowledges kind hospitality of the PSM, Dresden.

\appendix

\begin{figure}[tb]
\begin{centering}
\includegraphics[width=0.6\columnwidth]{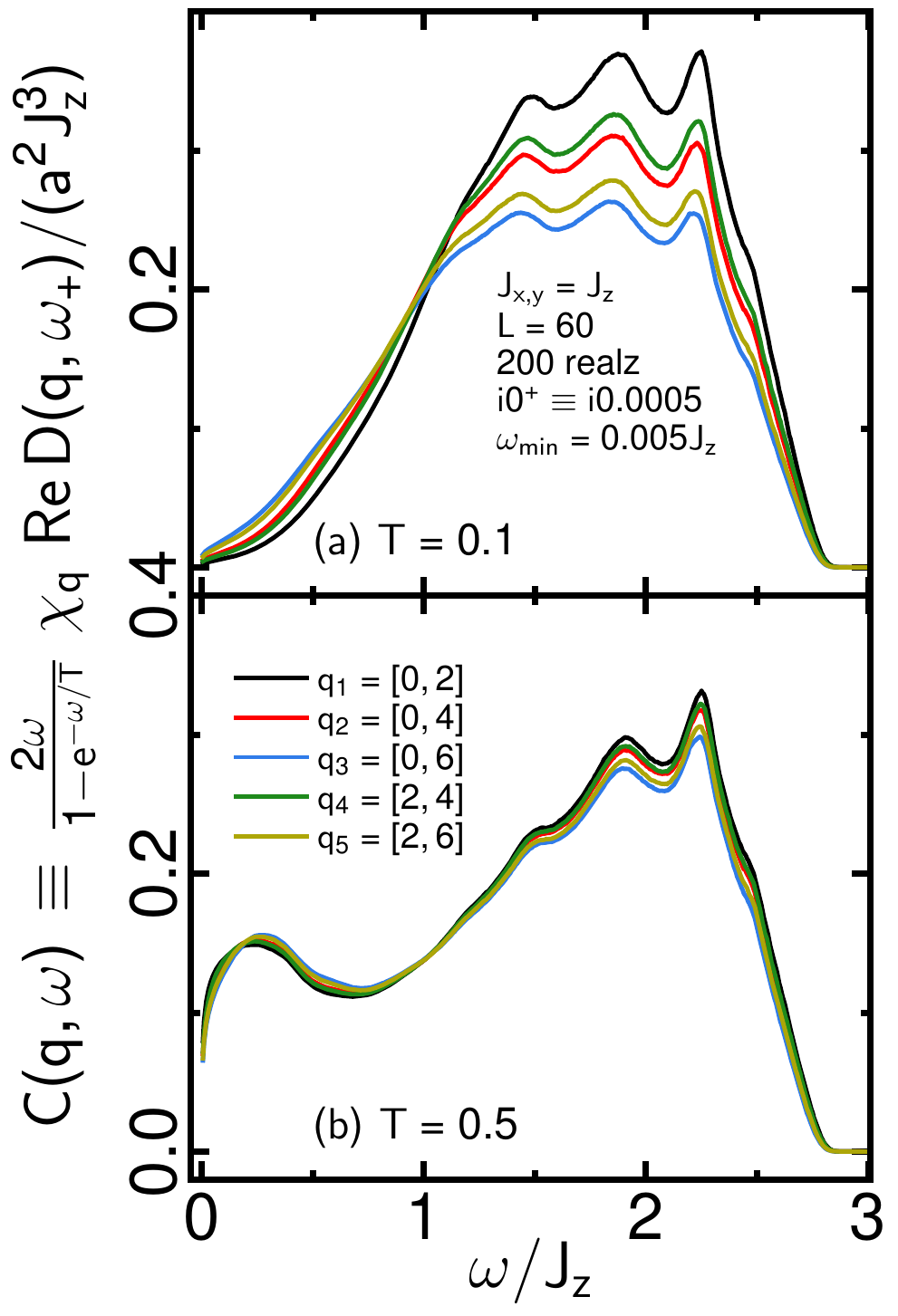} 
\end{centering}
\caption{\label{fig:8}Momentum evolution of Einstein relation for current
correlation function in the random gauge state, with 200 realizations, versus
$\omega\ge 0.005J_{z}$ for $T/J_{z}=0.1,$ and 0.5 in panels (a) and (b), for
various small momenta ${\bf q}=2\pi/L\sum_{j=1,2}q_{ij}{\bf G}_{j}$ within
the small-$q$ region identical to Fig. \ref{fig:5}, for $L^{2}=60\times60$
sites and for imaginary broadening $i0^{+}$.}
\end{figure}

\section{Current correlation functions}\label{app1}

In the limit of $\boldsymbol{q}\rightarrow\boldsymbol{0}$ one may
speculate, that the dynamical energy-density diffusion kernel is related
to the dynamical energy-current correlation function via a generalized
Einstein relation. The zero momentum current correlation function
has been considered in Ref.~\cite{Metavitsiadis2017}. For completeness
sake, we now clarify a relation of the latter quantity to the present
work. From Mori-Zwanzig's memory function method we have 
\begin{equation}
\frac{1}{z}(\chi_{{\bf q}} - \chi({\bf q},z))
=\frac{1}{z-M({\bf q},z)
\frac{1}{\chi_{{\bf q}}}}\chi_{{\bf q}}\,,
\label{eq:a1}
\end{equation}
where $M({\bf q}, z) = \langle Lh_{{\bf q}} | (z - QL)^{-1} QL h_{{\bf
q}}\rangle$ is the memory function. $L$ is the Liouville operator $LA=[H,A]$
and $\langle A|B\rangle=\int_{0}^{\beta}\langle A^{+}(\lambda)B\rangle
d\lambda-\beta\langle A^{+}\rangle\langle B\rangle$ is Mori's scalar product
with $A(\lambda)=e^{\lambda H}Ae^{-\lambda H}=e^{\lambda L}A$ and $\beta=1/T$
is the inverse temperature. $\chi_{{\bf q}}$ is the isothermal energy
susceptibility $\chi_{{\bf q}}=\langle h_{{\bf q}}|h_{{\bf q}}\rangle$ and
$Q$ is a projector perpendicular to the energy density, which is formulated
using Mori's product as $Q=1-|h_{{\bf q}}\rangle\chi_{{\bf q}}^{-1}\langle
h_{{\bf q}}|$.  We emphasize, that Eq. (\ref{eq:a1}) is \emph{not} a
``high-frequency'', or ``slow-mode'' approximation. It is a rigorous
statement. Due to time-reversal invariance, $QLh_{{\bf q}}=Lh_{{\bf q}}$
\cite{Forster1975}. Moreover, using the continuity equation in the
hydrodynamic regime, i.e., discarding the lattice structure, we have $Lh_{{\bf
q}}=-{\bf q}\cdot{\bf j_{{\bf q}}}$, where ${\bf j_{{\bf q}}}$ is the energy
current. Altogether
\begin{equation}
i\chi_{{\bf q}}D({\bf q},z) =
\sum_{\mu\nu}e_{\boldsymbol{q}\mu}
e_{\boldsymbol{q}\nu}\langle
j_{\boldsymbol{q}\mu}|\frac{1}{z-QL}j_{\boldsymbol{q}\nu}\rangle\,,
\label{eq:a2}
\end{equation}
where $\boldsymbol{e}_{\boldsymbol{q}\mu}$ are the components of the unit
vector into $\boldsymbol{q}$-direction. While for arbitrary $\boldsymbol{q}$
the right hand side refers to a so-called current relaxation-function with a
dynamics governed by a \emph{projected} Liouville operator $QL$, for
$\boldsymbol{q}\rightarrow\boldsymbol{0}$, one finds that
$\lim_{\boldsymbol{q}\rightarrow\boldsymbol{0}}\langle
j_{\boldsymbol{q}\mu}|(z-QL)^{-1}j_{\boldsymbol{q}\nu}\rangle=\langle
j_{\boldsymbol{0}\mu}|(z-L)^{-1}j_{\boldsymbol{0}\nu}\rangle$
\cite{Forster1975}, which is the genuine current relaxation-function
comprising the \emph{complete} Liouvillean dynamics. This turns Eq.
(\ref{eq:a2}) into an Einstein relation for $\boldsymbol{q} \rightarrow
\boldsymbol{0}$.  Finally, the spectrum of the current relaxation-function
can be related to that of a standard current correlation-function
$C_{\mu\nu}(t)=\langle j_{\boldsymbol{0}\mu}(t)j_{\boldsymbol{0}\nu}\rangle$
by the Kubo relation and the fluctuation dissipation theorem
\begin{equation}
\frac{2\omega}{1-e^{\omega/T}}\lim_{\boldsymbol{q}
\rightarrow\boldsymbol{0}}\chi_{{\bf q}}\,
\mathrm{Re}D({\bf q},z)=C(\omega)\,.
\label{eq:a3}
\end{equation}
Here we have discarded questions of anisotropy. While the present work's
focus is on $\boldsymbol{q}\neq0$, it is now very tempting to evaluate the
left hand side of Eq. (\ref{eq:a3}) using, e.g., the two temperatures of
Fig. \ref{fig:5} (a,b) and to consider its evolution with momentum. This is
shown in Fig. \ref{fig:8}, which should be compared with Fig. 5 (b,d) of
Ref. \cite{Metavitsiadis2017}. For this, and because of a different energy
unit and normalization of spectral densities in the latter Ref., $T$ has to
rescaled by $4$ and the y-axis by $4^{3}/\pi$. While the rescaled
temperatures $T=0.25$ and $0.525$ of Ref. \cite{Metavitsiadis2017} are not
absolutely identical to the ones we use, it is very satisfying to realize,
that the limit $\boldsymbol{q}\rightarrow\boldsymbol{0}$ of
Eq. (\ref{eq:a3}), which can be anticipated from Fig. \ref{fig:8} is
completely in line with Fig. 5 of Ref. \cite{Metavitsiadis2017}, including
the dip at very low $\omega$. This is even more remarkable in view of the
numerical representation and treatment of the Majorana fermions used in the
present work and in Ref. \cite{Metavitsiadis2017} being decisively different.

%\vspace*{1cm}

\end{document}